\documentclass[widetext,twocolumn,showpacs,preprintnumbers,superscriptaddress,nofootinbib]{revtex4}

\usepackage[dvipdfmx]{graphicx}
\usepackage{amssymb}
\usepackage{amsmath}
\usepackage{dcolumn}
\usepackage{bm}
\usepackage{natbib}
\usepackage{multirow}
\usepackage{subfigure}
\usepackage{url}

\makeatletter
\@addtoreset{equation}{section}

\makeatother

\begin{document}

\preprint{DESY 15-250}

\title{Axion dark matter in the post-inflationary Peccei-Quinn symmetry breaking scenario}

\author{Andreas Ringwald}
\affiliation{Deutsches Elektronen-Synchrotron DESY,
Notkestrasse 85, 22607 Hamburg, Germany} 

\author{Ken'ichi Saikawa}
\affiliation{Deutsches Elektronen-Synchrotron DESY,
Notkestrasse 85, 22607 Hamburg, Germany}
\affiliation{Department of Physics, Tokyo Institute of Technology,
2-12-1 Ookayama, Meguro-ku, Tokyo 152-8551, Japan}

\begin{abstract}
We consider extensions of the Standard Model in which a spontaneously broken global chiral Peccei-Quinn (PQ) symmetry arises as an accidental symmetry of an exact $Z_N$ symmetry. For $N = 9$ or $10$, this symmetry 
can protect the accion -- the Nambu-Goldstone boson arising from the spontaneous breaking of the accidental PQ symmetry -- 
against semi-classical gravity effects, thus suppressing gravitational corrections to the effective potential, while it can at the same time provide for the small explicit symmetry 
breaking term needed to make models with domain wall number $N_{\rm DW}>1$, such as the popular 
Dine-Fischler-Srednicki-Zhitnitsky (DFSZ) model ($N_{\rm DW}=6$), 
cosmologically viable even in the case where spontaneous PQ symmetry breaking occurred after inflation. 
We find that $N=10$ DFSZ accions with mass $m_A\approx 3.5 \textendash 4.2\,\mathrm{meV}$ can account for cold dark matter and simultaneously explain
the hints for anomalous cooling of white dwarfs.
The proposed helioscope International Axion Observatory -- being sensitive to solar DFSZ accions 
with mass above a few meV -- will decisively test this scenario.
\end{abstract}


\maketitle

\section{\label{sec1} Introduction}
The nature of dark matter is one of the greatest puzzles in particle physics and cosmology. 
One of the best motivated candidates is the axion $A$~\cite{Weinberg:1977ma,Wilczek:1977pj}. 
It arises 
as a pseudo Nambu-Goldstone (NG) boson from the spontaneous breaking of a hypothetical global chiral 
U(1)$_{\rm PQ}$ extension of the Standard Model (SM) which is  introduced to provide a solution 
to the strong CP problem~\cite{Peccei:1977hh}.  Soon after its introduction, it was realized 
that for large U(1)$_{\rm PQ}$ Peccei-Quinn (PQ) symmetry breaking scale, $v_{\rm PQ}\gg 10^{9}$\,GeV,  
the axion is also a cold dark matter candidate~\cite{Preskill:1982cy,Abbott:1982af,Dine:1982ah}. 

The prediction of the axion dark matter abundance depends strongly on the early history of the Universe, 
in particular on the fate of the U(1)$_{\rm PQ}$ symmetry during and after inflation (for recent reviews, 
see Refs.~\cite{Sikivie:2006ni,Ringwald:2012hr,Kawasaki:2013ae,Marsh:2015xka}). 
In this paper, we concentrate on the case that the reheating temperature of the Universe was 
high enough, $T_R> T_c$, such that the U(1)$_{\rm PQ}$ symmetry was restored
during reheating and spontaneously broken only later, when the temperature fell below the critical temperature of the PQ phase transition, $T_c\sim v_{\rm PQ}$. 
In this case, axion dark matter is produced not only by the re-alignment mechanism~\cite{Preskill:1982cy,Abbott:1982af,Dine:1982ah} but also by the  
decay of topological defects which has to be quantitatively taken into account~\cite{Davis:1986xc,Lyth:1991bb}.

In fact, vortex-like defects -- strings -- are formed at the PQ phase transition. Later, when the temperature of the Universe becomes comparable to the QCD phase transition, these strings are attached by surface-like defects -- domain walls.
The structure of the domain walls is determined by an integer number $N_{\rm DW}$ --  the domain wall number --  which  is related to the chiral U(1)$_{\rm PQ}$-SU(3)$_c$-SU(3)$_c$ anomaly. 

The evolution of these hybrid networks of strings and domain walls -- string-wall systems -- 
crucially depends on $N_{\rm DW}$. It is known that they are short-lived for $N_{\rm DW} = 1$, while they 
are long-lived for  $N_{\rm DW} > 1$. In fact, strictly speaking, for $N_{\rm DW} > 1$, domain walls are stable, 
as long as the U(1)$_{\rm PQ}$ is an exact global symmetry. In this case, they constitute a cosmological problem -- the domain wall problem -- since they would overclose the Universe, in conflict with standard 
cosmology~\cite{Zeldovich:1974uw,Sikivie:1982qv}.

One of the solutions to this problem is to introduce a small explicit U(1)$_{\rm PQ}$ symmetry breaking parameter in the Lagrangian, 
which makes the walls unstable and leads to their annihilation at late times~\cite{Gelmini:1988sf,Larsson:1996sp}. 
In fact, it is generically expected that the PQ symmetry -- like any global symmetry -- is not protected  
from explicit symmetry breaking effects by Planck-scale suppressed operators appearing in the low-energy effective Lagrangian~\cite{Georgi:1981pu,Ghigna:1992iv,Barr:1992qq,Kamionkowski:1992mf,Holman:1992us,Dine:1992vx,Rai:1992xw,Kallosh:1995hi,Dobrescu:1996jp}.
However, these operators modify also the axion potential, eventually shifting its minimum away from zero, thereby destroying the solution of the strong CP problem. 

Crucially, this drawback is absent in the models where the Peccei-Quinn symmetry is not {\it ad hoc}
but instead an automatic or accidental symmetry of an exact discrete $Z_N$ symmetry~\cite{Georgi:1981pu}. In fact, 
for $N \geq 9$, the discrete symmetry  
can protect the axion against semi-classical gravity effects~\cite{Dias:2002hz,Dias:2002gg,Dias:2003zt,Carpenter:2009zs,Harigaya:2013vja,Dias:2014osa}, 
while it can at the same time provide for the small explicit symmetry breaking term needed to make the 
$N_{\rm DW}>1$ models cosmologically viable.
In this case there still exists a pseudo NG boson arising from the spontaneous breaking of the accidental U(1)$_{\rm PQ}$ symmetry,
and we call it the {\it accion}~\cite{Choi:2009jt}.

The organization of this paper is as follows. In Sec.~\ref{sec2} we are considering Kim-Shifman-Vainshtein-Zakharov (KSVZ) like~\cite{Kim:1979if,Shifman:1979if} and 
Dine-Fischler-Srednicki-Zhitnitsky (DFSZ) like~\cite{Dine:1981rt,Zhitnitsky:1980tq} extensions of the SM which feature exact $Z_N$ symmetries.
We determine the lowest dimensional Planck-suppressed operators explicitly violating the PQ symmetry and 
discuss the parameter space available for the solution of the strong CP problem. 
In Sec.~\ref{sec3} we apply the results of Refs.~\cite{Hiramatsu:2010yu,Hiramatsu:2012gg,Hiramatsu:2012sc,Kawasaki:2014sqa}
on the prediction of axion dark matter in the post-inflationary PQ symmetry breaking scenario and translate them 
in the parameter space of our accion models. The prospects of the detection of accion dark matter are 
discussed in Sec.~\ref{sec4}. Finally, our conclusions and discussion are in Sec.~\ref{sec5}. 

\section{\label{sec2} Axion from accidental Peccei-Quinn symmetry}
Although the PQ mechanism with global U(1)$_{\rm PQ}$ symmetry dynamically solves the strong CP problem, such a global symmetry might not exist in nature
since it is likely to be violated by gravitational effects. Instead of introducing such {\it ad hoc} global symmetry, we consider the case where
the PQ symmetry emerges as an accidental symmetry from an exact discrete symmetry.
Various explicit examples have been considered in Ref.~\cite{Dias:2014osa}.

In this section, we consider two benchmark models with accidental PQ symmetry. We construct KSVZ-type models in Sec.~\ref{sec2-1} and
DFSZ-type models in Sec.~\ref{sec2-2}. We also discuss the constraint on the Planck-suppressed operators to avoid CP violation in Sec.~\ref{sec2-3}.

\begin{table}[ht]\centering
\begin{equation}{\nonumber
\begin{array}{|c|c|c|c|c|c|c|c|c|c|}
\hline\rule[0cm]{0cm}{.9em}
&q_L & u_R & d_R & L & l_R & H & Q_L & Q_R  &  \sigma
\\[-.1ex]
\hline\rule[0cm]{0cm}{1em}
Z_{9} & 1 & \omega_{9}  & \omega^8_{9} & 1 & \omega^8_{9} & \omega_9 &
1 & \omega_{9}  &  \omega^8_{9}
\\[-.1ex]
\hline\rule[0cm]{0cm}{1em}
Z_{10} & 1 & \omega^6_{10} & \omega^4_{10} & 1 & \omega^4_{10} & \omega^6_{10} &
\omega^5_{10} & \omega^6_{10}  &  \omega^9_{10}
\\[-.1ex] \hline
\end{array}
}
\end{equation}
\caption{The $Z_N$ charges (for $N=9,10$), where $\omega_N\equiv e^{i2\pi/N}$,  of the KSVZ accion models, leaving the Yukawa interactions \eqref{lyuk_ksvz_I},~\eqref{lyuk_ksvz_II} and~\eqref{lyuk_ksvz_III} invariant.}
\label{tab:KSVZ_discrete}
\end{table}

\subsection{\label{sec2-1} KSVZ accion models}
Like in the KSVZ model~\cite{Kim:1979if,Shifman:1979if}, we extend the field content of the SM by a complex SM singlet scalar field $\sigma$ and 
a color-triplet exotic quark $Q$. 
We assume an exact discrete $Z_N$ symmetry in which the fields transform according to Table~\ref{tab:KSVZ_discrete}.
With the assumption that $Q$ is an SU(2)$_L\times$U(1)$_Y$ singlet (``KSVZ I"), the Yukawa interaction terms consistent with the charge assignments in Table~\ref{tab:KSVZ_discrete}
are given by
\begin{align}
{\mathcal L_Y} & =   Y_{ij}\overline{q}_{iL}\widetilde{H} u_{j R}  +\Gamma_{ij}\overline{q}_{iL} H d_{j R} + G_{ij}\overline{L}_{i } H l_{j R}\nonumber\\
&\quad+ y_Q \overline{Q}_L \sigma Q_R + h.c.\,,
\label{lyuk_ksvz_I}
\end{align}
where $H$ is the SM Higgs doublet; $\widetilde{H}=\epsilon H^*$;
$Y_{ij}$, $\Gamma_{ij}$ and $G_{ij}$ are complex
$3\times 3$ matrices; and $i,j=1,2,3$ are flavor indices.
Note that other renormalizable operators which are invariant under the SM gauge symmetries
are forbidden because of the exact discrete symmetry,
except for hermitian terms in the scalar potential, which lead to the spontaneous breaking of the PQ and the electroweak symmetry.
The Yukawa interactions shown in Eq.~\eqref{lyuk_ksvz_I} are invariant under an accidental U(1)$_{\rm PQ}$
symmetry with the charge assignments shown in the first line of Table~\ref{tab:KSVZ_PQ} labeled ``KSVZ I". 

\begin{table}
$$
\begin{array}{|l|ccccccccc|}
\hline\rule[0cm]{0cm}{1em}
 \mathrm{Model} &q_L & u_R & d_R & L & l_R & H   & Q_L & Q_R & \sigma \\[.3ex]
 \hline\rule[0cm]{0cm}{1em}
\mathrm{KSVZ\ I} & 0 & 0 & 0 & 0 & 0 & 0  & 1/2 & -1/2 & 1 
\\[.5ex]
\hline\rule[0cm]{0cm}{1em}
\mathrm{KSVZ\ II} & 3/2 & 3/2 & 3/2 & 0 & 0 & 0  & 1/2 & -1/2 & 1 
\\[.5ex]
\hline\rule[0cm]{0cm}{1em}
\mathrm{KSVZ\ III} & -1/2 & -1/2 & -1/2 & 0 & 0 & 0  & 1/2 & -1/2 & 1 
\\[.5ex]
\hline
\end{array}
$$
\caption{The U(1)$_{\rm PQ}$ charge assignments leaving \eqref{lyuk_ksvz_I} (KSVZ I), \eqref{lyuk_ksvz_I} plus \eqref{lyuk_ksvz_II} (KSVZ II) or \eqref{lyuk_ksvz_I} plus \eqref{lyuk_ksvz_III} (KSVZ III) invariant.}
\label{tab:KSVZ_PQ}
\end{table}

We assume that the scalar interaction potential is such that the singlet scalar $\sigma$ acquires a vacuum expectation value (VEV) $\langle |\sigma|^2\rangle=v^2_{\rm PQ}/2$. In the broken phase of the accidental PQ symmetry,  
the accion NG field $A$ is identified as the phase direction of the scalar field $\sigma(x)\propto e^{iA(x)/v_{\rm PQ}}$.

The color-triplet $Q$ may also be charged under the electroweak symmetry. Here we consider the following three possibilities according to the hypercharge $Y_{Q_R}$ of $Q_R$,
\begin{equation}
Y_{Q_R} = 
\left\{
\begin{array}{rl}
0 & (\mathrm{KSVZ\ I})\\
-\frac{1}{3} & (\mathrm{KSVZ\ II})\\
\frac{2}{3} & (\mathrm{KSVZ\ III}).
\end{array}
\right.
\end{equation}
The KSVZ I model corresponds to the Yukawa interactions given by Eq.~\eqref{lyuk_ksvz_I}. 
In the case where $Q_R$ has a non-zero hypercharge, there are additional dimension 4 operators,
which are invariant under $Z_N$ and the SM gauge symmetry, e.g.,
\begin{equation}
\bar{Q}_L\sigma^* d_R,
\label{lyuk_ksvz_II}
\end{equation}
for KSVZ II, and
\begin{equation}
\bar{Q}_L\sigma u_R,\qquad \bar{q}_L\tilde{H}Q_R,
\label{lyuk_ksvz_III}
\end{equation}
for KSVZ III.\footnote{If we assume different charge assignments of the discrete symmetry,
other operators such as $\bar{Q}_L\sigma d_R$, $\bar{q}_L HQ_R$, and $\bar{Q}_L\sigma^*u_R$
would appear in the Lagrangian. This ambiguity does not affect the main conclusion of the paper.}
Then, the total Lagrangian is invariant under the PQ symmetry transformations whose charge assignments are shown in Table~\ref{tab:KSVZ_PQ}.

In the KSVZ I model, the exotic quarks $Q$ decay into SM particles only through higher dimensional operators, which lead to a long lifetime.  
The existence of such long-lived particles is cosmologically problematic and may conflict with several observational results~\cite{Nardi:1990ku,Berezhiani:1992rk}.
This problem can be avoided in KSVZ II and III, since exotic quarks rapidly decay into lighter particles through renormalizable interactions given by Eq.~\eqref{lyuk_ksvz_II}
or Eq.~\eqref{lyuk_ksvz_III}. 

\begin{table}[ht]\centering
\begin{equation}{\nonumber
\begin{array}{|l|c|c|c|c|c|c|c|c|}
\hline\rule[0cm]{0cm}{.9em}
&q_L & u_R & d_R & L & l_R & H_u & H_d  &  \sigma
\\[-.1ex]
\hline\rule[0cm]{0cm}{1em}
Z_{9}\ (\mathrm{DFSZ\ I}) & 1 & \omega^6_{9}  & \omega^5_{9} & 1 & \omega^5_{9} & \omega^6_{9} & \omega^4_{9} & \omega_{9}
\\[-.1ex]
\hline\rule[0cm]{0cm}{1em}
Z_{9}\ (\mathrm{DFSZ\ II}) & 1 & \omega^4_{9}  & \omega^7_{9} & 1 & \omega^5_{9} & \omega^4_{9} & \omega^2_{9} & \omega_{9}
\\[-.1ex]
\hline\rule[0cm]{0cm}{1em}
Z_{10}\ (\mathrm{DFSZ\ I}) & 1 & \omega^3_{10}  & \omega^9_{10} & 1 & \omega^9_{10} & \omega^3_{10} & \omega_{10} & \omega_{10}
\\[-.1ex]
\hline\rule[0cm]{0cm}{1em}
Z_{10}\ (\mathrm{DFSZ\ II}) & 1 & \omega^3_{10}  & \omega^9_{10} & 1 & \omega^7_{10} & \omega^3_{10} & \omega_{10} & \omega_{10}
\\[-.1ex] \hline
\end{array}
}
\end{equation}
\caption{The $Z_N$ charges (for $N=9,10$) of the DFSZ accion models, leaving interactions \eqref{lyuk_dfsz_I} and \eqref{vnh_dfsz} (DFSZ I) or \eqref{lyuk_dfsz_II} and \eqref{vnh_dfsz} (DFSZ II) invariant.}
\label{tab:DFSZ_discrete}
\end{table}

\subsection{\label{sec2-2} DFSZ accion models}
Like in the DFSZ model \cite{Dine:1981rt,Zhitnitsky:1980tq}, we extend the SM field content by the familiar complex singlet scalar $\sigma$ and by introducing two Higgs doublets $H_u$ and $H_d$, whose VEVs give masses
to up-type quarks and down-type quarks, respectively.
There are two possibilities according to whether leptons couple to $H_d$ (DFSZ I) or $H_u$ (DFSZ II). 
We assume that the theory possesses a discrete $Z_N$ symmetry with the charge assignments as shown in Table~\ref{tab:DFSZ_discrete}. Then, the Yukawa interaction terms
consistent with the discrete $Z_N$ symmetry read
\begin{align}
\label{lyuk_dfsz_I}
{\mathcal L_Y} = \Gamma_{ij}\overline{q}_{iL} H_d d_{j R}  
+ Y_{ij}\overline{q}_{iL}\widetilde{H}_u u_{j R}
+ G_{ij}\overline{L}_{i } H_d l_{j R} + h.c.\,,
\end{align}
for DFSZ I, and
\begin{align}
\label{lyuk_dfsz_II}
{\mathcal L_Y} = \Gamma_{ij}\overline{q}_{iL} H_d d_{j R}  
+ Y_{ij}\overline{q}_{iL}\widetilde{H}_u u_{j R}
+ G_{ij}\overline{L}_{i } H_u l_{j R} + h.c.\,,
\end{align}
for DFSZ II. There also exist the non-hermitian terms in the scalar potential,
\begin{equation}
\label{vnh_dfsz}
V_{NH} = \lambda H_d^{\dagger}H_u\sigma^{*2} + h.c.\,,
\end{equation}
where $\lambda$ is a dimensionless constant.
Other gauge invariant renormalizable operators are forbidden because of the exact discrete symmetry,
except for hermitian terms in the scalar potential, which lead to the spontaneous breaking of the PQ and the electroweak symmetry.
The interactions given by Eqs.~\eqref{lyuk_dfsz_I} and~\eqref{vnh_dfsz} (DFSZ I) or Eqs.~\eqref{lyuk_dfsz_II} and~\eqref{vnh_dfsz} (DFSZ II)
are invariant under an accidental U(1)$_{\rm PQ}$ symmetry with the charge assignments shown in Table~\ref{tab:DFSZ_PQ}.

\begin{table}[h]
$$
\begin{array}{|l|cccccccc|}
\hline\rule[0cm]{0cm}{1em}
 \mathrm{Model} &q_L & u_R & d_R & L & l_R & H _u & H_d & \sigma \\[.3ex]
 \hline\rule[0cm]{0cm}{1em}
\mathrm{DFSZ\ I} & 0 & X_u & X_d & 0 & X_d & X_u & -X_d & 1 
\\[.5ex]
\hline\rule[0cm]{0cm}{1em}
\mathrm{DFSZ\ II} & 0 & X_u & X_d & 0 & -X_u & X_u & -X_d & 1 
\\[.5ex]
\hline
\end{array}
$$
\caption{The U(1)$_{\rm PQ}$ charge assignments, where $X_u$ and $X_d$ are some real numbers satisfying the condition $X_u+X_d=2$, leaving \eqref{lyuk_dfsz_I} and \eqref{vnh_dfsz} (DFSZ I) or \eqref{lyuk_dfsz_II} and \eqref{vnh_dfsz} (DFSZ II) invariant.}
\label{tab:DFSZ_PQ}
\end{table}

After the PQ and the electroweak phase transition, the singlet field and the neutral components of the Higgs doublets acquire VEVs, 
$\langle\sigma\rangle = v_{\rm PQ}/\sqrt{2}$, $\langle H_d^0\rangle = v_d/\sqrt{2}$ and $\langle H_u^0\rangle = v_u/\sqrt{2}$.
The accion field $A(x)$ can be parametrized in terms of phase directions of three scalar fields, i.e.,
$H_d^0(x) \propto e^{-iX_d A(x)/\tilde{f}_A}$, $H_u^0(x) \propto e^{iX_u A(x)/\tilde{f}_A}$, and $\sigma(x) \propto e^{iA(x)/\tilde{f}_A}$.
The orthogonality of $A(x)$ and the NG boson eaten by the  $Z^0$ boson implies $X_d=x^{-1}\xi_v$ and $X_u=x\xi_v$, where\footnote{The prime at the angle
indicates that we chose the convention usually taken in the axion literature~\cite{Srednicki:1985xd}.
It is related to the convention in the Higgs literature by $\tan\beta \equiv v_u/v_d = \cot\beta'$.}
\begin{equation}
x \equiv \frac{v_d}{v_u} \equiv \tan\beta'. 
\end{equation}
From the condition $X_u+X_d=2$, we obtain $\xi_v=2/(x+x^{-1})$.
Furthermore, requiring the canonical normalization of $A(x)$ in the low-energy effective Lagrangian, we obtain the relation
\begin{equation}
\tilde{f}_A = \sqrt{v_{\rm PQ}^2+v^2\xi_v^2},
\end{equation}
where $v=\sqrt{v_u^2+v_d^2} \simeq 246\,\mathrm{GeV}$ is the SM Higgs VEV.

\subsection{\label{sec2-3} Constraint from CP violation}
The global U(1)$_{\rm PQ}$ symmetry of the accion models is explicitly broken due to the existence of 
higher dimensional Planck-suppressed operators.
In particular, there exist operators of the form
\begin{align}
{\mathcal L} \supset
\frac{g}{M_{\rm P}^{D-4}}\mathcal{O}_D=
\frac{g}{M_{\rm P}^{D-4}}\left(H^\dagger_b H_c\right)^m\sigma^l,
\label{Planck_sup_operators}
\end{align}
where $l$ and $m$ are integers, $D=2m+l > 4$ is the mass dimension of the operator, $H^\dagger_b H_c$ corresponds to $H^\dagger H$ in the KSVZ models and
$H^\dagger_u H_u$, $H^\dagger_d H_d$, $H^\dagger_u H_d$, or $H^\dagger_d H_u$ in the DFSZ models, 
$M_{\rm P} \simeq 2.435\times 10^{18}\,\mathrm{GeV}$ is the reduced Planck mass, and $g$ is a dimensionless constant.
Note that, in general, $g$ is a complex number.
The operator shown in Eq.~\eqref{Planck_sup_operators} can be consistent with the discrete symmetry but break the U(1)$_{\rm PQ}$ symmetry, which leads to a CP violation.
The lowest dimensional operators that are suppressed by $M_{\rm P}$ and break the PQ symmetry read
\begin{equation}
\mathcal{O}_D = \left\{
\begin{array}{ll}
\sigma^N & \mathrm{for\ KSVZ\ models} \\
 (H_d^{\dagger}H_u)^m\sigma^{N-2m} &  \mathrm{for\ DFSZ\ models},
\end{array}
\right.
\end{equation}
with $D=N$ for $Z_N$ symmetry.
If $\langle\sigma\rangle\gg \langle H_u^0\rangle,\langle H_d^0\rangle$, the largest contribution is $\mathcal{O}_D = \sigma^N$ for both the KSVZ and DFSZ models.
Hereafter, we only consider this $\sigma^N$ term and ignore the effects from other higher dimensional operators.

In the presence of the higher dimensional operators, the effective potential for the accion field reads~\cite{Dias:2014osa,DiVecchia:1980yfw,diCortona:2015ldu}
\begin{align}
V_{\rm eff} &\simeq -m_{\pi}^2 f_{\pi}^2\sqrt{1-\frac{4m_u m_d}{(m_u+m_d)^2}\sin^2\left(\frac{A}{2f_A}\right)} \nonumber\\
&\quad- \frac{|g|v_{\rm PQ}^N}{(\sqrt{2})^{N-2}M_{\rm P}^{N-4}}\cos\left(N\frac{A}{\tilde{f}_A} + \Delta_D\right) \label{V_eff_accions}
\end{align}
for $Z_N$ symmetry, where $m_\pi\simeq 135\,\mathrm{MeV}$ is the pion mass, $f_\pi \simeq 92\,\mathrm{MeV}$
is the pion decay constant, $m_u$ and $m_d$ are up and down quark mass, respectively, and
\begin{equation}
\Delta_D = \Delta - N\bar{\theta}.
\end{equation}
Here, $\Delta$ is the phase of the coupling $g$ (i.e. $g=|g|e^{i\Delta}$), and $\bar{\theta}$ consists of
the QCD $\theta$ parameter and the contribution from the phase of the quark masses.

The first term on the right-hand side of Eq.~\eqref{V_eff_accions} arises due to topological fluctuations of the gluon fields in QCD~\cite{DiVecchia:1980yfw,diCortona:2015ldu}. 
If we can ignore the second term on the right-hand side of Eq.~\eqref{V_eff_accions},
the effective potential has a minimum at $\langle A\rangle/f_A=0$, which dynamically solves the strong CP problem.
Expanding this potential to the quadratic order in $A/f_A$, we can obtain the expression for the accion mass,
\begin{align}
m_A = \frac{m_\pi f_\pi}{f_A}\frac{\sqrt{z}}{1+z} \simeq 6\times 10^{-6}\mathrm{eV}\left(\frac{10^{12}\,\mathrm{GeV}}{f_A}\right),
\end{align}
where $z=m_u/m_d=0.38$-$0.58$ is the ratio of up and down quark masses, and
$f_A$ is the accion decay constant, which satisfies the relation
\begin{equation}
f_A = \frac{\tilde{f}_A}{N_{\rm DW}},
\end{equation}
where $\tilde{f}_A$ is given by
\begin{equation}
\tilde{f}_A = \left\{
\begin{array}{ll}
v_{\rm PQ} & \mathrm{for\ KSVZ\ models} \\
\sqrt{v_{\rm PQ}^2+v^2\xi_v^2} &  \mathrm{for\ DFSZ\ models}.
\end{array}
\right.
\end{equation}
The value of the domain wall number $N_{\rm DW}$ is determined by the color anomaly coefficient~\cite{Sikivie:1982qv,Georgi:1982ph,Kim:1986ax,Dias:2014osa},
\begin{equation}
N_{\rm DW} = \left\{
\begin{array}{ll}
1 & \mathrm{for\ KSVZ\ models} \\
2 N_g &  \mathrm{for\ DFSZ\ models},
\end{array}
\right.
\label{NDW_values}
\end{equation}
where $N_g=3$ is the number of generations.

Because of the existence of the second term on the right-hand side of Eq.~\eqref{V_eff_accions},
the minimum of the effective potential is shifted from the value $\langle A\rangle/f_A=0$.
Such a shift causes a CP violation, whose magnitude is severely constrained by the observation of the neutron electric dipole moment (NEDM)~\cite{Baker:2006ts}.
Expanding the right-hand side of Eq.~\eqref{V_eff_accions} for small $A/f_A$, we obtain
\begin{align}
V_{\rm eff} &\simeq \frac{1}{2} m_A^2 A^2 \nonumber\\
&\quad+\frac{1}{2}\frac{N^2|g|N_{\rm DW}^{N-2}}{(\sqrt{2})^{N-2}}\left(\frac{f_A}{M_{\rm P}}\right)^{N-2}M_{\rm P}^2\cos\Delta_D A^2 \nonumber\\
&\quad+\frac{N|g|N_{\rm DW}^{N-1}}{(\sqrt{2})^{N-2}}\left(\frac{f_A}{M_{\rm P}}\right)^{N-1}M_{\rm P}^3\sin\Delta_D A,
\label{V_eff_accions_approx}
\end{align}
where we used $v_{\rm PQ}\simeq \tilde{f}_A = N_{\rm DW}f_A$.
In order to satisfy the experimental bound from the NEDM, we require~\cite{Kim:2008hd}
\begin{align}
\frac{|\langle A\rangle|}{f_A} &\simeq \frac{\dfrac{N|g|N_{\rm DW}^{N-1}}{(\sqrt{2})^{N-2}}\left(\dfrac{f_A}{M_{\rm P}}\right)^{N-2}M_{\rm P}^2\sin\Delta_D}{m_A^2 + \dfrac{N^2|g|N_{\rm DW}^{N-2}}{(\sqrt{2})^{N-2}}\left(\dfrac{f_A}{M_{\rm P}}\right)^{N-2}M_{\rm P}^2\cos\Delta_D}\nonumber\\
& < 0.7\times 10^{-11}.  \label{constraint_NEDM}
\end{align}

If the order $N$ of the discrete symmetry is small enough, the coefficient of the $A^2$ term in Eq.~\eqref{V_eff_accions_approx}
is dominated by the term proportional to $|g|$. In this case the accion mass is determined by the higher dimension operators rather than the QCD induced potential.
In order to avoid it, we require the following condition:
\begin{equation}
m_g^2 \equiv \frac{N^2|g|N_{\rm DW}^{N-2}}{(\sqrt{2})^{N-2}}\left(\frac{f_A}{M_{\rm P}}\right)^{N-2}M_{\rm P}^2\cos\Delta_D < m_A^2.
\label{condition_m_g}
\end{equation}
For instance, if $N=8$, we obtain
\begin{align}
m_g &\simeq 1.03\times 10^2 \mathrm{eV}\ |g|^{1/2}\left(\cos\Delta_D\right)^{1/2}\nonumber\\
&\quad\times\left(\frac{N_{\rm DW}}{6}\right)^3\left(\frac{f_A}{10^9\,\mathrm{GeV}}\right)^3.
\end{align}
This cannot be smaller than $m_A$ for $f_A > \mathcal{O}(10^8\textendash10^9)\mathrm{GeV}$ unless $|g|$ is extremely small.
The constraint becomes more severe if we take a smaller value for $N$.
This fact implies that $N\ge 9$ is preferable from the viewpoint of the strong CP problem.

\section{\label{sec3} Accion dark matter from topological defects}
Let us estimate now the relic abundance of accion dark matter in the post-inflationary PQ symmetry breaking scenario by using the results of numerical simulations
obtained in Refs.~\cite{Hiramatsu:2010yu,Hiramatsu:2012gg,Hiramatsu:2012sc,Kawasaki:2014sqa}.
If the PQ symmetry is broken after inflation, there are three contributions: accions produced via the re-alignment mechanism, those produced by strings,
and those produced by the annihilation of the string-wall systems.
Strings are formed around the epoch of the PQ phase transition due to the spontaneous breaking of the global U(1)$_{\rm PQ}$ symmetry,
while string-wall systems are formed around the epoch of the QCD phase transition due to the spontaneous breaking of the $Z_{N_{\rm DW}}$ symmetry, which is a subgroup of the U(1)$_{\rm PQ}$.

The estimation for the contribution from string-wall systems differs depending on whether $N_{\rm DW}=1$ or $N_{\rm DW}>1$.
If $N_{\rm DW}=1$, the string-wall systems are unstable, and they start to collapse immediately after their formation at the epoch of the QCD phase transition.
In this case, the contribution from the string-wall systems turns out to be comparable with those from other production mechanisms~\cite{Hiramatsu:2012gg}.
On the other hand, if $N_{\rm DW}>1$, a string is attached by $N_{\rm DW}$ domain walls.
Such a configuration can exist for a long time, and it only collapses at a later time due to the effect of the explicit symmetry breaking term in the potential.
In this case, the contribution from the long-lived string-wall systems can be much larger than those from other production mechanisms~\cite{Hiramatsu:2012sc}.

From Eq.~\eqref{NDW_values}, we see that the string-wall systems are short-lived for the KSVZ models ($N_{\rm DW}=1$)
and long-lived for the DFSZ models ($N_{\rm DW}=2 N_g >1$).
Therefore, there is a significant difference on the prediction for the dark matter abundance between the KSVZ-type and DFSZ-type models.
We discuss its observational consequences in Sec.~\ref{sec4}.

Before going to the estimation of the contribution from string-wall systems, let us briefly summarize the results for the two
further contributions.
The contribution from the re-alignment mechanism is given by~\cite{Kawasaki:2014sqa}
\begin{align}
\Omega_{A,\mathrm{real}}h^2 &= 
4.63\times10^{-3} \nonumber\\
&\quad\times 
\left(\frac{f_A}{10^{10}\,\mathrm{GeV}}\right)^{(6+n)/(4+n)}\left(\frac{\Lambda_{\rm QCD}}{400\,\mathrm{MeV}}\right),
\label{Omega_A_mis}
\end{align}
where the exponent $n=6.68$ arises from the non-trivial temperature dependence of the accion mass~\cite{Wantz:2009it},
and $\Lambda_{\rm QCD}\approx 400\,\mathrm{MeV}$ is 
a fit parameter of the order of the QCD scale.\footnote{In Ref.~\cite{Wantz:2009it}, the temperature dependence was taken from
the instanton liquid model. A re-evaluation based on QCD lattice calculations of the topological susceptibility 
along the lines of Refs.~\cite{Borsanyi:2015cka,Bonati:2015vqz} is under way.}
In order to estimate the contribution from the strings, we must follow the evolution of
the global strings from the time of the PQ phase transition to that of the QCD phase transition.
According to the results of numerical simulations~\cite{Yamaguchi:1998gx,Hiramatsu:2010yu,Kawasaki:2014sqa}, 
accions produced by global strings have a mean energy comparable to the Hubble parameter at the production time,
which leads to the following estimation for the relic abundance:
\begin{align}
\Omega_{A,\mathrm{string}}h^2 & \simeq (7.3\pm3.9)\times 10^{-3}\times N_{\rm DW}^2 \nonumber\\
&\quad\times\left(\frac{f_A}{10^{10}\,\mathrm{GeV}}\right)^{(6+n)/(4+n)}\left(\frac{\Lambda_{\rm QCD}}{400\,\mathrm{MeV}}\right).
\label{Omega_A_string}
\end{align}

The estimation of the contribution from the string-wall systems is model dependent.
First, we consider the KSVZ-type models with $N_{\rm DW}=1.$
The collapse of the short-lived string-wall systems in these models was investigated in Ref.~\cite{Hiramatsu:2012gg}.
It turns out that the mean energy of the radiated accions is comparable to their mass and that the contribution from
the string-wall systems can be estimated as~\cite{Kawasaki:2014sqa}
\begin{align}
\Omega_{A,\mathrm{wall}}h^2 &= (3.7\pm1.4)\times 10^{-3}\nonumber\\
&\quad\times
\left(\frac{f_A}{10^{10}\,\mathrm{GeV}}\right)^{(6+n)/(4+n)}\left(\frac{\Lambda_{\rm QCD}}{400\,\mathrm{MeV}}\right).
\label{Omega_A_wall_short}
\end{align}
The total accion dark matter abundance is given by the sum of the three contributions:
\begin{equation}
\Omega_{A,\mathrm{tot}}h^2 = \Omega_{A,\mathrm{real}}h^2 + \Omega_{A,\mathrm{string}}h^2 + \Omega_{A,\mathrm{wall}}h^2. \label{Omega_A_tot}
\end{equation}
Assuming that $\Omega_{A,\mathrm{tot}}h^2$ accounts for the total cold dark matter abundance observed today, $\Omega_{\rm CDM}h^2 \simeq 0.12$~\cite{Ade:2013zuv},
we obtain the following prediction for the decay constant in the KSVZ models,
\begin{equation}
f_A \approx (4.6 \textendash 7.2)\times 10^{10}\,\mathrm{GeV}, \label{fA_prediction_KSVZ}
\end{equation}
which corresponds to the accion mass 
$m_A \approx (0.8 \textendash 1.3) \times 10^{-4}\,\mathrm{eV}$.

Next, we consider the DFSZ-type models with $N_{\rm DW}=2N_g=6$.
The energy density of the long-lived string-wall systems can be approximated by that of the domain walls.
According to the recent numerical simulation, the time evolution of the energy density of the domain walls can be modeled as~\cite{Kawasaki:2014sqa}
\begin{equation}
\rho_{\rm wall}(t) = \frac{\mathcal{A}(t)\sigma_{\rm wall}}{t}\quad\mathrm{with}\quad \mathcal{A}(t) = \mathcal{A}_{\rm form}\left(\frac{t}{t_{\rm form}}\right)^{1-p},
\end{equation}
where $\sigma_{\rm wall} \simeq 9.23 m_A f_A^2$ is the surface mass density of domain walls, $t_{\rm form}$ is the time for the formation of them, and the coefficient $\mathcal{A}_{\rm form}$ is measured in the simulations.
We may naively expect that the energy density obeys a simple scaling relation $\rho_{\rm wall}(t)\sim\sigma_{\rm wall}/t$ from dimensional analysis, but the results of numerical simulations show some deviation from the exact scaling behavior ($p=1$). Taking account of this uncertainty,
we consider two possibilities: exact scaling ($p=1$) and deviation from scaling ($p\ne 1$).

At a later time, the collapse of the domain walls occurs due to the existence of the Planck suppressed operator in Eq.~\eqref{V_eff_accions},
which can be parametrized as
\begin{equation}
V_{\rm eff} \supset -2\Xi v_{\rm PQ}^4\cos\left(N\frac{A}{\tilde{f}_A}+\Delta_D\right),
\end{equation}
where
\begin{equation}
\Xi \equiv \frac{|g|}{(\sqrt{2})^N}\left(\frac{v_{\rm PQ}}{M_{\rm P}}\right)^{N-4} \simeq \frac{|g|N_{\rm DW}^{N-4}}{(\sqrt{2})^N}\left(\frac{f_A}{M_{\rm P}}\right)^{N-4}.
\label{def_Xi}
\end{equation}
Assuming that the field value at the domain having the lowest energy is given by $\langle A\rangle/\tilde{f}_A\simeq 0$,
we estimate the energy difference between the lowest energy domain and its neighbor $\langle A\rangle/\tilde{f}_A\simeq 2\pi/N_{\rm DW}$ as\footnote{We note that 
if $\Delta_D\ll1$, Eq.~\eqref{DeltaVeff} can be approximated as $\Delta V_{\rm eff}\simeq 2\Xi v_{\rm PQ}^4(1-\cos(2\pi N/N_{\rm DW}))$,
which results in a factor $(1-\cos(2\pi N/N_{\rm DW}))$ appearing in Eqs.~\eqref{t_dec},~\eqref{Omega_A_wall_exact} and~\eqref{Omega_A_wall_dev}~\cite{Kawasaki:2014sqa}.
Strictly speaking, this approximation does not hold for $\Delta_D\simeq\mathcal{O}(1)$, and 
we must use the full expression [Eq.~\eqref{DeltaVeff}] rather than the approximated form.
However, this difference does not change the final result for the dark matter abundance within the error margin.}
\begin{align}
\Delta V_{\rm eff} &\simeq -2\Xi v_{\rm PQ}^4\left[\cos\left(\frac{2\pi N}{N_{\rm DW}}+\Delta_D\right)-\cos\Delta_D\right].
\label{DeltaVeff}
\end{align}

The energy difference shown in Eq.~\eqref{DeltaVeff} acts as a volume pressure $p_V\sim \Delta V_{\rm eff}$ on domain walls.
Their collapse occurs when this pressure effect becomes comparable to the tension of domain walls $p_T\sim \mathcal{A}\sigma_{\rm wall}/t$.
From the relation $p_V\sim p_T$, we estimate the decay time of domain walls:
\begin{equation}
t_{\rm dec} = C_d\left[\frac{\mathcal{A}_{\rm form}\sigma_{\rm wall}}{t_{\rm form}\Xi v_{\rm PQ}^4(1-\cos(2\pi N/N_{\rm DW}))}\right]^{1/p}t_{\rm form},
\label{t_dec}
\end{equation}
where the coefficient $C_d$ is determined from numerical simulations.

The long-lived domain walls copiously produce accions until the decay time $t_{\rm dec}$. The mean energy of the radiated accions is given by~\cite{Kawasaki:2014sqa}
\begin{equation}
\bar{\omega}_A = \tilde{\epsilon}_A m_A,
\end{equation}
where $\tilde{\epsilon}_A$ is a coefficient of $\mathcal{O}(1)$, whose value can be determined from numerical simulations.
Using the outcomes of the simulations ($\tilde{\epsilon}_A$, $\mathcal{A}$, $p$, $C_d$), we can estimate the present
abundance of accions produced by domain walls~\cite{Kawasaki:2014sqa}:
\begin{align}
\Omega_{A,\mathrm{wall}}h^2 &= 0.756\times\frac{C_d^{1/2}}{\tilde{\epsilon}_A}\left[\frac{\mathcal{A}^3}{N_{\rm DW}^4(1-\cos(2\pi N/N_{\rm DW}))}\right]^{1/2}\nonumber\\
&\quad\times\left(\frac{\Xi}{10^{-52}}\right)^{-1/2}\left(\frac{f_A}{10^{10}\,\mathrm{GeV}}\right)^{-1/2}\nonumber\\
&\quad\times\left(\frac{\Lambda_{\rm QCD}}{400\,\mathrm{MeV}}\right)^3,
\label{Omega_A_wall_exact}
\end{align}
for the assumption of exact scaling ($p=1$), and
\begin{align}
\Omega_{A,\mathrm{wall}}h^2
&= 1.23\times 10^{-6}\times[7.22\times 10^3]^{3/2p}\times\frac{1}{\tilde{\epsilon}_A}\frac{2p-1}{3-2p}\nonumber\\
&\quad\times C_d^{3/2-p}\mathcal{A}_{\rm form}^{3/2p}\nonumber\\
&\quad\times \left[N_{\rm DW}^4\left(1-\cos\left(\frac{2\pi N}{N_{\rm DW}}\right)\right)\right]^{1-3/2p} \nonumber\\
&\quad\times
\left(\frac{\Xi}{10^{-52}}\right)^{1-3/2p}\left(\frac{f_A}{10^{10}\,\mathrm{GeV}}\right)^{4+\frac{3(4p-16-3n)}{2p(4+n)}}\nonumber\\
&\quad\times\left(\frac{\Lambda_{\rm QCD}}{400\,\mathrm{MeV}}\right)^{-3+6/p},
\label{Omega_A_wall_dev}
\end{align}
for the assumption of deviation from scaling ($p\ne1$).
The total accion abundance is again given by the sum~\eqref{Omega_A_tot}.
It is notable that in the DFSZ models $\Omega_{A,\mathrm{wall}}h^2$ can be much larger than the other two contributions
$\Omega_{A,\mathrm{real}}h^2$ and $\Omega_{A,\mathrm{string}}h^2$ for small values of $f_A$.

Note that for $N=11$ the value of $\Xi$ in Eq.~\eqref{def_Xi} becomes
\begin{equation}
\Xi = 1.22\times 10^{-55}\times |g|\left(\frac{N_{\rm DW}}{6}\right)^7\left(\frac{f_A}{10^{10}\,\mathrm{GeV}}\right)^7.
\end{equation}
Substituting it into Eq.~\eqref{Omega_A_wall_exact}, we obtain $\Omega_{A,\mathrm{wall}}h^2\gtrsim 0.26$
for $|g|=1$. Obviously, it violates the observational constraint $\Omega_{A,\mathrm{tot}}h^2 \lesssim 0.12$.
Therefore, we must require $N\le 10$ in order to avoid the overproduction of accion dark matter from the domain walls.
Recall that there is another requirement, $N\ge 9$, which follows from Eq.~\eqref{condition_m_g}.
Combining these two facts, we conclude that the order of the discrete symmetry for the phenomenologically viable model is $N=9$ or $10$.

\begin{figure}[htbp]
\centering
$\begin{array}{c}
\subfigure[]{
\includegraphics[width=0.45\textwidth]{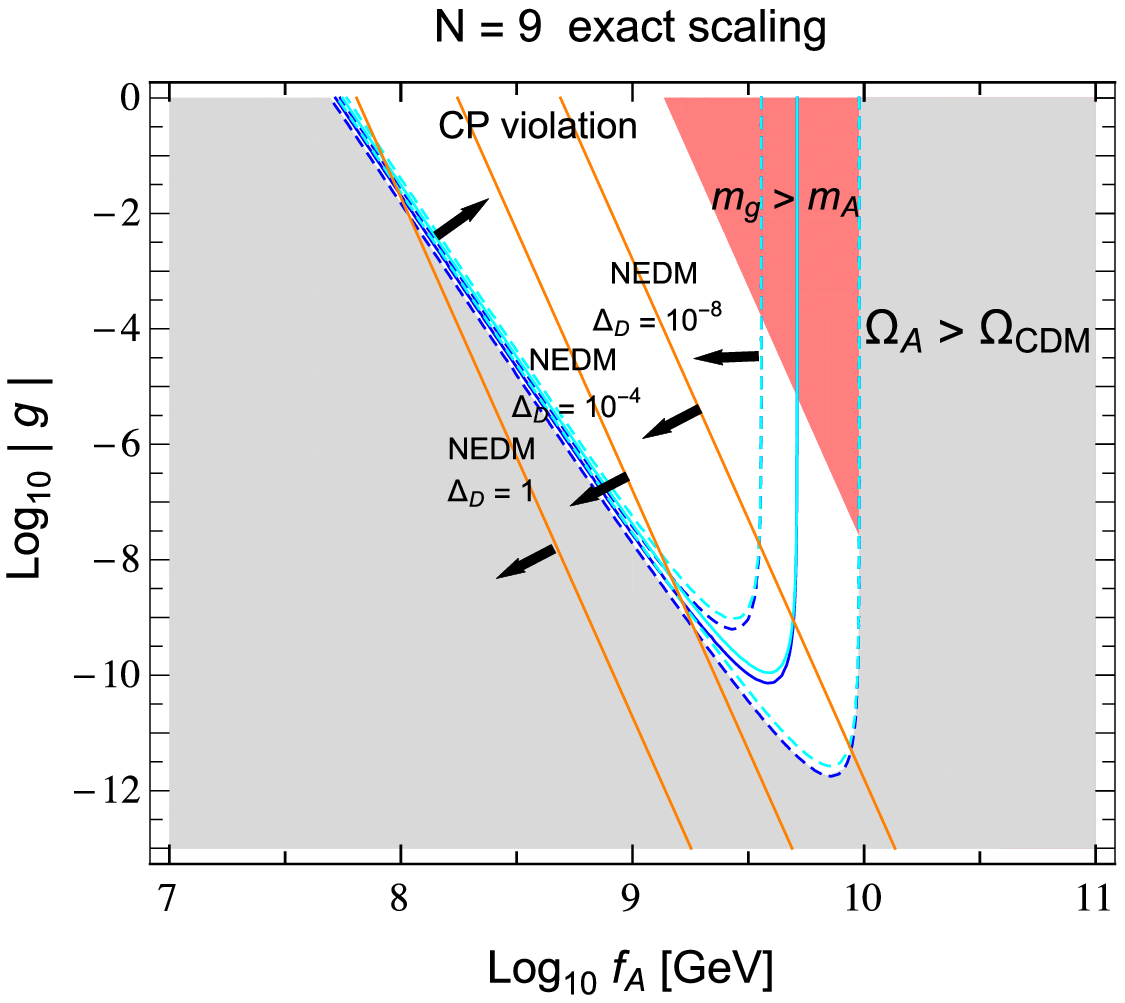}}
\\\\
\subfigure[]{
\includegraphics[width=0.45\textwidth]{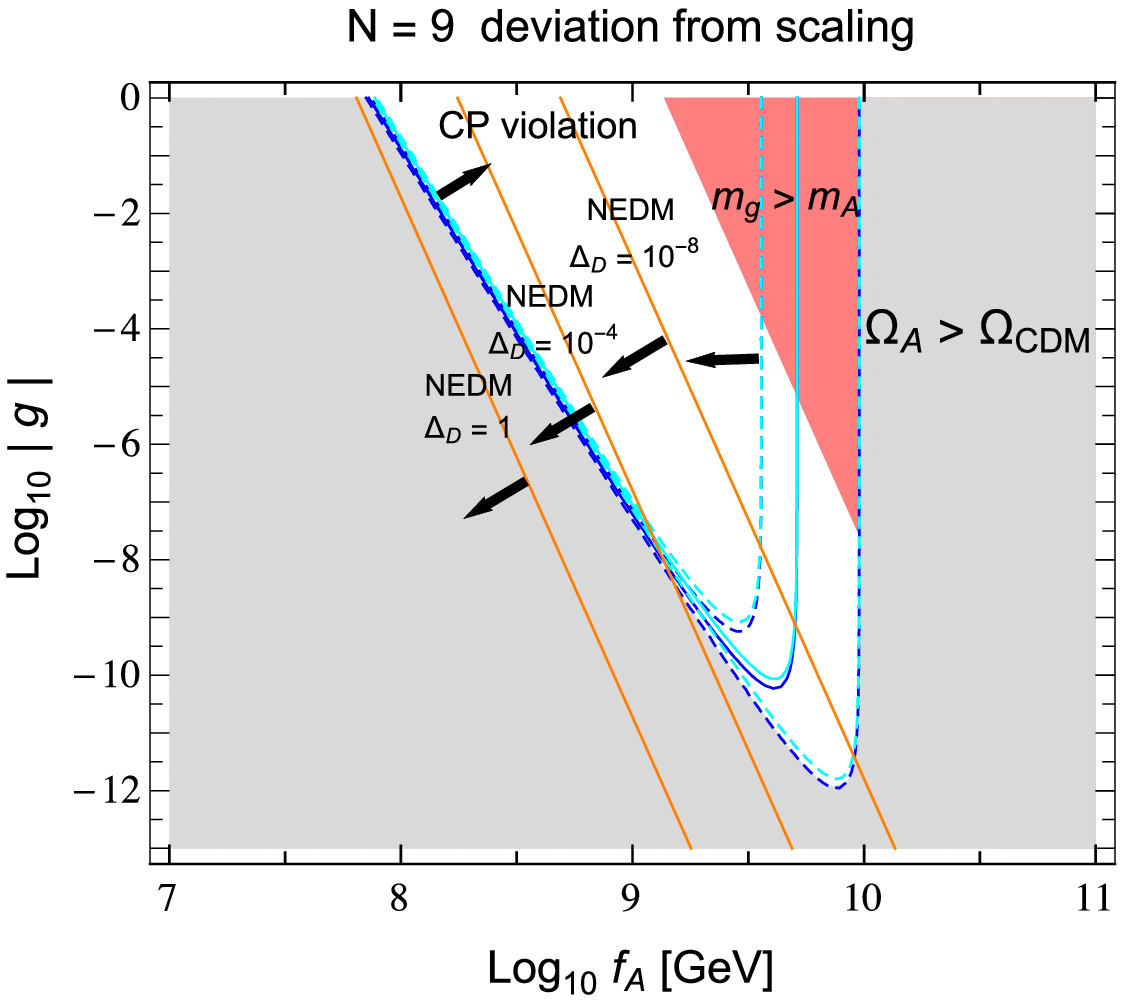}}
\end{array}$
\caption{Plot of the observational constraints on the parameter space of $(f_A,|g|)$ in the DFSZ accion models
for the assumption of exact scaling [panel (a)] and that of deviation from scaling [panel (b)].
Here, we take $N=9$ for the discrete symmetry and $N_{\rm DW}=6$ for the domain wall number.
The region below the blue or cyan lines is excluded since the accion abundance exceeds the observed cold dark matter abundance $\Omega_{\rm CDM}h^2\simeq 0.12$.
On the blue (cyan) lines, the lifetime of the domain walls are estimated based on the criterion that the area density of the walls becomes smaller than $10\%$ ($1\%$) of that with $\Xi=0$.
The dotted blue (cyan) lines represent uncertainties of the estimation of the relic accion abundance.
In the red region, the condition given by Eq.~\eqref{condition_m_g} is violated, and the Planck-suppressed operator dominates over the QCD potential for the accion field.
The yellow lines correspond to the bound from the CP violation [Eq.~\eqref{constraint_NEDM}] for $\Delta_D=1$, $10^{-4}$, and $10^{-8}$.
The region above these lines is also excluded according to the value of $\Delta_D$.}
\label{fig:f_A-g_plot_N9}
\end{figure}

\begin{figure}[htbp]
\centering
$\begin{array}{c}
\subfigure[]{
\includegraphics[width=0.45\textwidth]{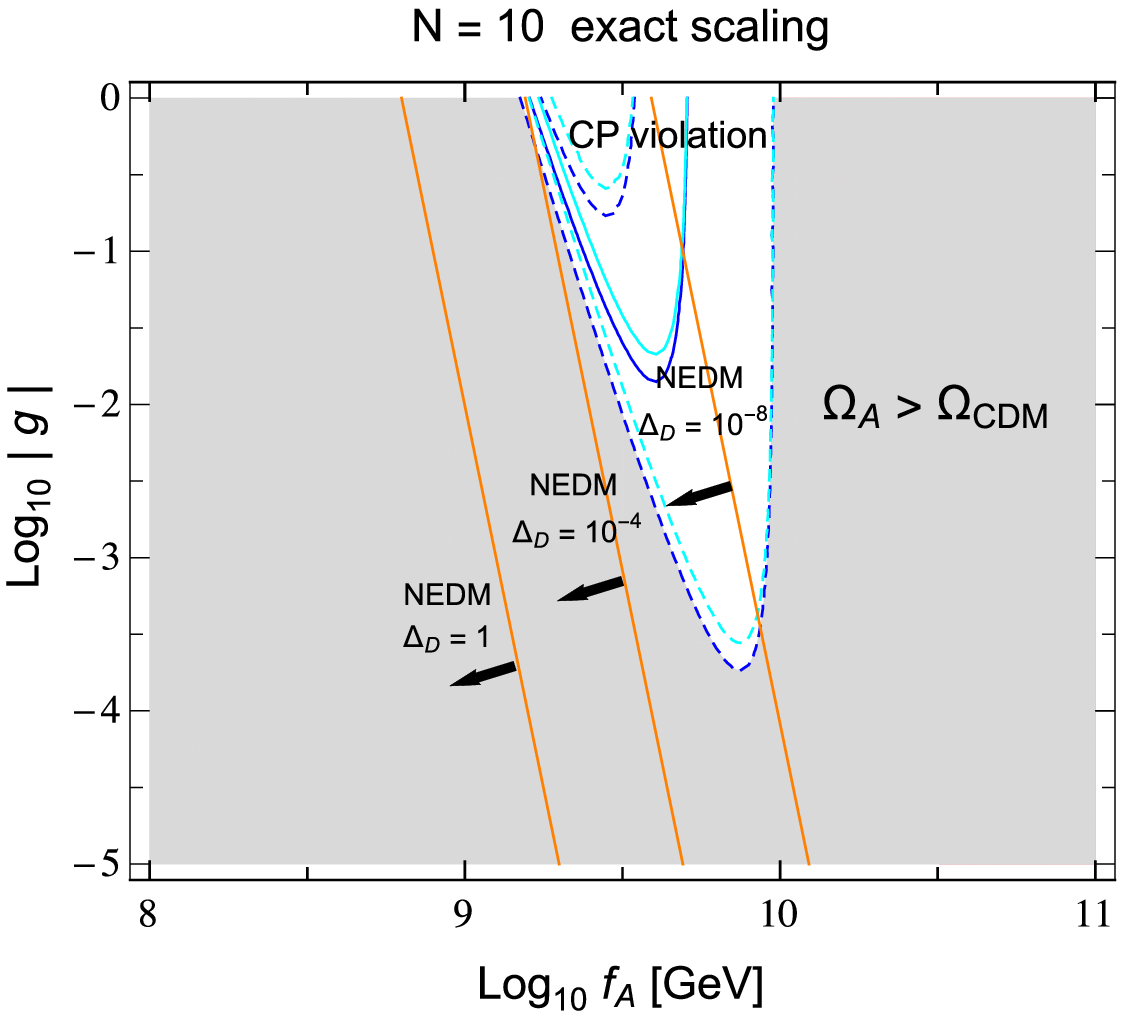}}
\\\\
\subfigure[]{
\includegraphics[width=0.45\textwidth]{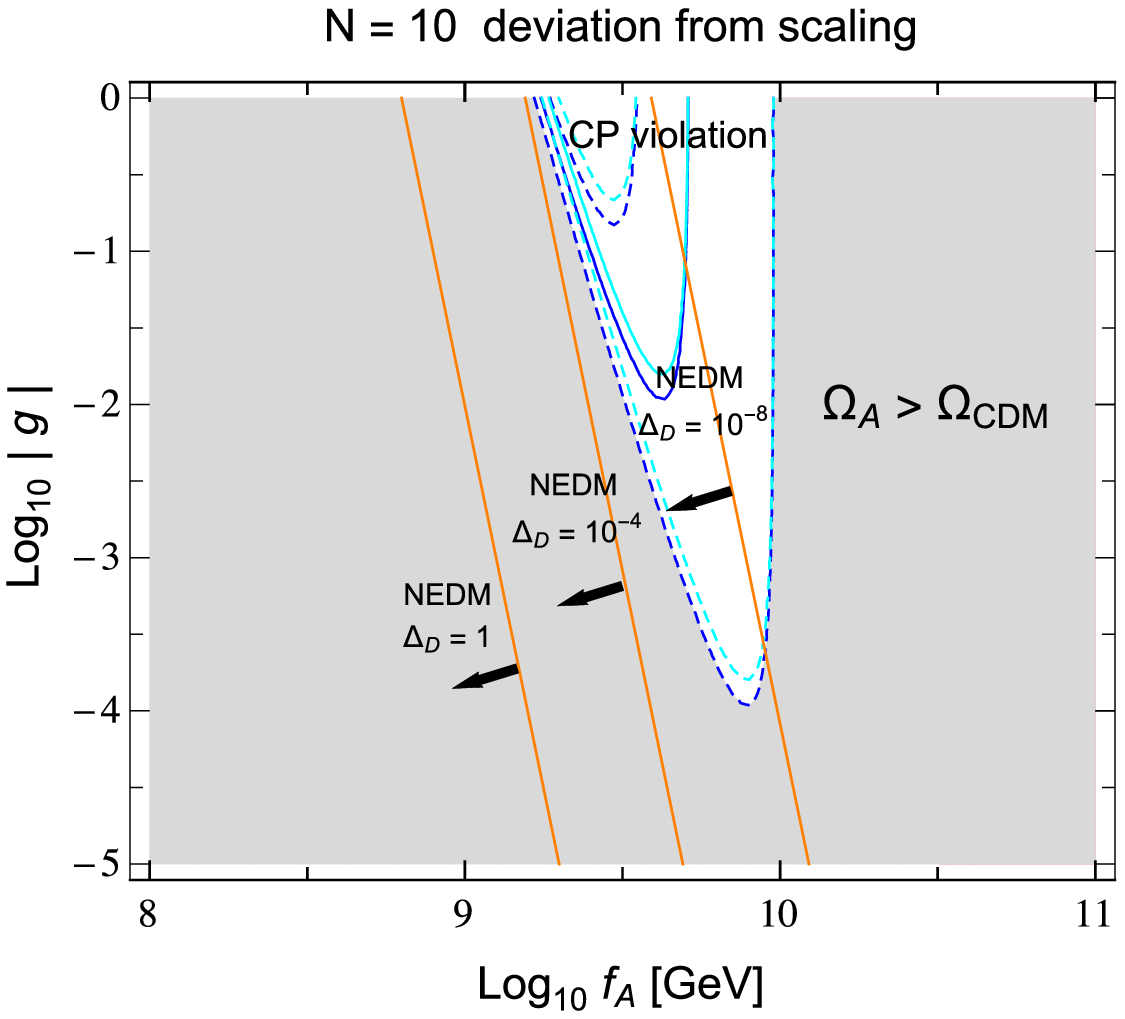}}
\end{array}$
\caption{The same figure as Fig.~\ref{fig:f_A-g_plot_N9}, but observational
constraints are plotted for the DFSZ accion models with $N=10$ based
on (a) the assumption of exact scaling and (b) that of deviation from scaling.}
\label{fig:f_A-g_plot_N10}
\end{figure}

We show the constraint in the $f_A$-$|g|$ plane for the case with $N=9$ in Fig.~\ref{fig:f_A-g_plot_N9} and $N=10$ in Fig~\ref{fig:f_A-g_plot_N10}.
We see that the dark matter constraint $\Omega_{A,\mathrm{tot}}h^2<\Omega_{\rm CDM}h^2$ gives a lower bound on $|g|$, while the NEDM constraint [Eq.~\eqref{constraint_NEDM}]
gives an upper bound on $|g|$.
The NEDM bound is relaxed if we allow a tuning in the phase parameter $\Delta_D$.
For instance, for the case with $N=10$ (Fig.~\ref{fig:f_A-g_plot_N10}) the whole parameter region is excluded if $\Delta_D\simeq \mathcal{O}(1)$, but the allowed region appears if $\Delta_D \lesssim \mathcal{O}(10^{-5}\textendash10^{-4})$.


On the blue (cyan) line in Figs.~\ref{fig:f_A-g_plot_N9} and~\ref{fig:f_A-g_plot_N10}, DFSZ accions can explain the observed dark matter abundance.
This corresponds to 
\begin{equation}
f_A \approx (5.5 \textendash 8.6)\times 10^7\,\mathrm{GeV}
\end{equation}
or $m_A \approx (0.7 \textendash 1.1) \times 10^{-1}\,\mathrm{eV}$ for $|g|=1$ and $N=9$ (see Fig.~\ref{fig:f_A-g_plot_N9}).
For $|g|=1$ and $N=10$ (Fig.~\ref{fig:f_A-g_plot_N10}), there are two parameter regions in which accions become the dominant component of dark matter:
\begin{equation}
f_A \approx (1.4 \textendash 1.7)\times 10^9\,\mathrm{GeV},
\end{equation}
which corresponds to $m_A \approx (3.5 \textendash 4.2) \times 10^{-3}\,\mathrm{eV}$, and 
\begin{equation}
f_A \approx (3.6 \textendash 9.5)\times 10^9\,\mathrm{GeV},
\end{equation}
which corresponds to $m_A \approx (0.6 \textendash 1.7) \times 10^{-3}\,\mathrm{eV}$.
The former requires a tuning $\Delta_D \lesssim \mathcal{O}(10^{-5}\textendash10^{-4})$, and in this parameter region, the contribution from the string-wall systems dominates over other contributions.
The latter requires a more severe tuning of $\Delta_D$, and in this parameter region, the contribution from the string-wall systems becomes irrelevant.
If we allow further tunings of the parameters $|g|$ and $\Delta_D$, the observed dark matter abundance can be explained in wider ranges of $f_A$.
In any case, the predicted values for $f_A$ are different from that obtained in the KSVZ models [Eq.~\eqref{fA_prediction_KSVZ}].
This is because the string contribution is enhanced by the factor of $N_{\rm DW}^2$ [see Eq.~\eqref{Omega_A_string}], and the contribution from the string-wall systems is further enhanced if they are long-lived.

\section{\label{sec4} Prospects to detect accion dark matter}
In this section, we discuss implications of the accion dark matter prediction for the present astrophysical observations and future experimental tests.
The various observations constrain or measure couplings of the accion to photons, nucleons, and charged leptons.
The interaction terms in the low-energy effective Lagrangian of $A$ at energies below the QCD scale are given by~\cite{Kaplan:1985dv,Srednicki:1985xd}
\begin{align}
\mathcal L_{\rm int} &= -\frac{\alpha}{8\pi}C_{A\gamma}\frac{A}{f_A}F_{\mu\nu}\tilde{F}^{\mu\nu} + \frac{1}{2}\sum_{N=p,n}C_{AN}\frac{\partial_{\mu}A}{f_A}\bar{\psi}_N\gamma^{\mu}\gamma\psi_N \nonumber\\
&\quad+\frac{1}{2}\sum_{\ell=e,\mu,\tau}C_{A\ell}\frac{\partial_{\mu}A}{f_A}\bar{\ell}\gamma^{\mu}\gamma \ell,
\label{accion_leff_int}
\end{align}
where $\alpha$ is the fine-structure constant and $\psi_N$ with $N=p$ and $n$ correspond to the proton field and the neutron field, respectively.
The nucleon couplings are given by~\cite{Raffelt:2006cw}
\begin{align}
\label{proton_coupling}
  C_{Ap}&=(C_{Au}-\eta)\Delta u
       +(C_{Ad}-\eta z)\Delta d
       +(C_{As}-\eta w)\Delta s\,,\\ 
\label{neutron_coupling}
  C_{An}&=(C_{Au}-\eta)\Delta d
       +(C_{Ad}-\eta z)\Delta u
       +(C_{As}-\eta w)\Delta s\,,
\end{align}
where $\eta=(1+z+w)^{-1}$ with $z=m_u/m_d=0.38\textendash0.58$ and $w=m_u/m_s$,
$\Delta u=0.84\pm0.02$, $\Delta d=-0.43\pm0.02$ and
$\Delta s=-0.09\pm0.02$.
The quark and lepton couplings ($C_{Au}$,$C_{Ad}$,$C_{As}$,$C_{A\ell}$)
are determined by their PQ charges, while the photon coupling $C_{A\gamma}$ contains a model-independent contribution arising from the mixing with pions.
We summarize the values of the dimensionless couplings $C_{A\gamma}$ and $C_{Ai}$ ($i=u,d,s,\ell$)
for the KSVZ and DFSZ accion models in Table \ref{tab:couplings}. 

\begin{table}
$$
\begin{array}{|l|ccccc|}
\hline
 \rm Model & C_{A\gamma} & C_{Au} & C_{Ad} &  C_{As} & C_{A\ell}   \\
\hline\hline
 \rm KSVZ\ I & -\frac{2}{3}\frac{4+z}{1+z} & 0 & 0  &  0 & 0  \\
\hline\hline
 \rm KSVZ\ II & \frac{2}{3}-\frac{2}{3}\frac{4+z}{1+z} & 0 & 0  &  0 & 0  \\
\hline\hline
 \rm KSVZ\ III & \frac{8}{3}-\frac{2}{3}\frac{4+z}{1+z} & 0 & 0  &  0 & 0  \\
\hline\hline
 \rm DFSZ\ I & \frac{8}{3}-\frac{2}{3}\frac{4+z}{1+z} & \frac{1}{3} \sin^2 \beta' & \frac{1}{3} \cos^2 \beta' & \frac{1}{3} \cos^2 \beta' & \frac{1}{3} \cos^2 \beta' \\
\hline\hline
 \rm DFSZ\ II &\frac{2}{3}-\frac{2}{3}\frac{4+z}{1+z} & \frac{1}{3} \sin^2 \beta' & \frac{1}{3} \cos^2 \beta' & \frac{1}{3} \cos^2 \beta' & -\frac{1}{3} \sin^2 \beta'
\\[.5ex]
\hline
\end{array}
$$
\caption{\label{tab:couplings}
Dimensionless couplings of the accion to SM particles. 
}
\end{table}

Various astrophysical observations have established stringent bounds on the couplings of axion-like particles (ALPs)
based on the energy-loss arguments due to the existence of ``dark channels" (i.e. ALPs). 
The recent analysis of the horizontal branch (HB) stars in galactic globular-clusters (GCs)~\cite{Ayala:2014pea}
finds an upper bound on the photon coupling,
\begin{equation}
|g_{A\gamma}| \equiv \frac{\alpha}{2\pi}\frac{|C_{A\gamma}|}{f_A} < 6.6\times 10^{-11}\mathrm{GeV}^{-1}\quad(95\%\ \mathrm{CL}).
\end{equation}
The high-precision photometry for the galactic globular cluster M5 (NGC 5904) provided in Ref.~\cite{Viaux:2013lha} allows us to investigate the red giant (RG) branch of the color-magnitude diagram
of GCs, which leads to an upper bound on the electron coupling,
\begin{equation}
\alpha_{Ae} \equiv \frac{g_{Ae}^2}{4\pi} \equiv \frac{C_{Ae}^2m_e^2}{4\pi f_A^2} < 1.5\times 10^{-26}\quad(95\%\ \mathrm{CL}).
\end{equation}
The electron coupling can also be constrained by using the luminosity function of white dwarfs (WDLF).
A recent detailed analysis~\cite{Bertolami:2014wua} finds that WDLFs disfavor the electron coupling above
\begin{equation}
\alpha_{Ae} \gtrsim 6\times 10^{-27}.
\end{equation} 
On the other hand, the fitting of some WDLFs is improved for the electron coupling in the range~\cite{Bertolami:2014wua,Isern:2008nt,Isern:2008fs}
\begin{equation}
4.1\times 10^{-28} \lesssim \alpha_{Ae} \lesssim 3.7\times 10^{-27}, \label{WD_cooling_hint}
\end{equation}
which might be interpreted as the existence of an accion in the meV mass range.
Finally, the observed duration of the neutrino burst from supernova SN1987A leads to a bound on nucleon couplings~\cite{Raffelt:2006cw}
\begin{equation}
g^2_{Ap}+2g^2_{An} \lesssim (9\times 10^{-10})^2,
\end{equation}
where
\begin{equation}
g_{AN} \equiv \frac{C_{AN}m_N}{f_A}.
\end{equation}

\begin{figure*}[htbp]
\begin{center}
\includegraphics[scale=1.0]{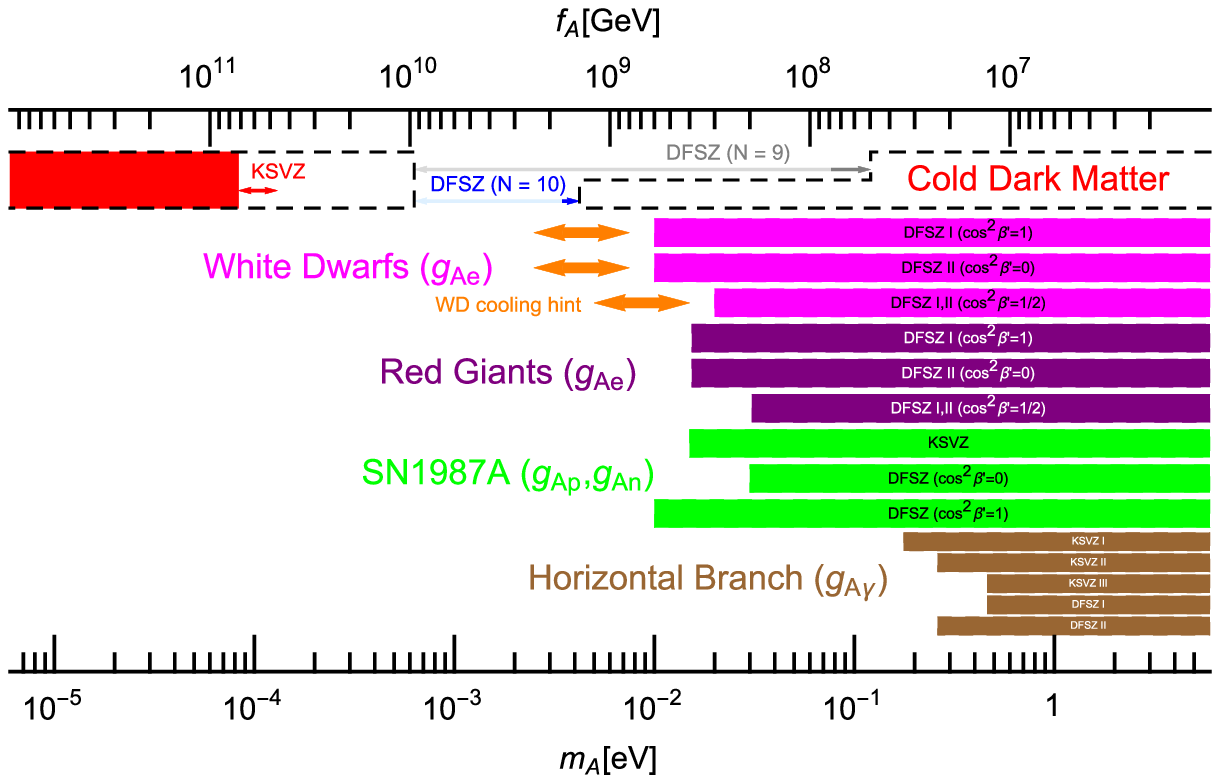}
\end{center}
\caption{Astrophysical and cosmological constraints on the accion decay constant $f_A$ or the accion mass $m_A$.
The colored intervals represent the bounds obtained by the cold dark matter overproduction in the KSVZ models (red),
WDLFs (magenta), RGs (purple), SN1987A (green), and HBs (brown).
In the DFSZ models, the region enclosed by the dashed lines is also excluded because of the cold dark matter overproduction.
The orange arrows represent the mass range favored by WDLFs [Eq.~\eqref{WD_cooling_hint}].
The red arrow represents the predicted mass range, where accions explain the observed cold dark matter abundance in the KSVZ models,
while the gray and blue arrows represent that in the DFSZ models with $N=9$ and $N=10$, respectively.
The deep gray (blue) line corresponds to the prediction obtained for the case with $|g|=1$ and a minimal tuning of $\Delta_D$, 
while the light gray (blue) line corresponds to that allowed for the case with $|g|<1$ and/or a further tuning of $\Delta_D$ in the $N=9$ $(10)$ DFSZ models.}
\label{fig:fAmA}
\end{figure*}

Figure~\ref{fig:fAmA} summarizes the astrophysical bounds and dark matter constraints on the KSVZ and DFSZ accion models.
The astrophysical observations give lower (upper) bounds on $f_A$ ($m_A$), while the dark matter abundance gives upper (lower) bounds on it.
The dark matter abundance also leads to a lower (upper) bound on $f_A$ ($m_A$) for the DFSZ models since in this region
the contribution from the string-wall systems becomes important and the dependence on $f_A$ [Eq.~\eqref{Omega_A_wall_exact} or~\eqref{Omega_A_wall_dev}] becomes different from 
that of the KSVZ models [Eqs.~\eqref{Omega_A_string},~\eqref{Omega_A_mis}, and~\eqref{Omega_A_wall_short}].
The bounds from RGs and WDLFs disappear for the KSVZ models since the electron coupling vanishes as shown in Table~\ref{tab:couplings}.
These bounds depend on $\tan\beta'$ in the DFSZ models, and they also disappear for the DFSZ I (II) accion for the case $\cos^2\beta'=0$ ($\cos^2\beta'=1$).
On the other hand, the bound from SN1987A becomes important for both the DFSZ I and DFSZ II accion models since
$C_{Ap}$ and $C_{An}$ cannot vanish simultaneously for any value of $\tan\beta'$.
From Fig.~\ref{fig:fAmA}, we see that the DFSZ models with $N=9$ and $|g|=1$ (gray line) are excluded from the SN1987A limit,
but there still exists the unexcluded parameter range for the $N=9$ DFSZ models (light gray line) if we allow smaller values of $|g|$ and a more severe tuning of $\Delta_D$.
We also note that the prediction of the DFSZ models with $N=10$ and $|g|=1$, where the contribution from the string-wall systems becomes dominant
with a tuning $\Delta_D \lesssim \mathcal{O}(10^{-5}\textendash10^{-4})$ (blue line), 
coincides with the parameter range suggested by the fitting of WDLFs.


The predicted mass ranges of accion dark matter have important implications for various ongoing and future experimental searches of ALPs.
Some lower mass ranges $m_A\gtrsim \mu\mathrm{eV}$ have already been probed by haloscope-type experiments~\cite{Sikivie:1983ip} 
using microwave cavity detectors such as the Axion Dark Matter Experiment (ADMX)~\cite{Asztalos:2001tf,Asztalos:2003px,Asztalos:2009yp} and earlier attempts~\cite{DePanfilis:1987dk,Wuensch:1989sa,Hagmann:1990tj}.
It is planned that higher mass ranges will be probed by using higher harmonic ports (ADMX-HF)~\cite{vanBibber:2013ssa}.
There are also helioscope-type experiments~\cite{Sikivie:1983ip} which observe ALPs produced by the sun. The CERN Axion Solar Telescope (CAST)~\cite{Arik:2013nya} has put a limit on
the photon coupling $g_{A\gamma}$ and on the combination $g_{A\gamma}\cdot g_{Ae}$~\cite{Barth:2013sma},
and a proposed helioscope, the International Axion Observatory (IAXO)~\cite{Armengaud:2014gea}, will probe a broad mass range with improved sensitivities.
Crucially, it can also probe the electron coupling $g_{Ae}$ in the range suggested by an meV mass DFSZ 
accion~\cite{Raffelt:2011ft,Armengaud:2013LoI}. 
The purely laboratory based light-shining-through-walls experiments~\cite{Redondo:2010dp} such as the Any Light Particle Search II (ALPS II)~\cite{Bahre:2013ywa},
which does not rely on astrophysical or cosmological assumptions, will also probe the photon coupling.

In addition to the experimental projects described above, there are several proposals to search an intermediate mass range of ALP dark matter.
The Orpheus experiment~\cite{Rybka:2014cya} uses an open Fabry-P\'erot resonator, which enables one to probe the photon coupling in the mass range $40\textendash400\,\mu\mathrm{eV}$.
It is also proposed to search for ALP dark matter by using atomic transitions~\cite{Sikivie:2014lha}, which will probe the electron coupling $g_{Ae}$ and
the nucleon couplings $g_{AN}$ in a similar mass range.
The electron coupling will also be probed by the Quaerere Axions (QUAX) experiment, which is proposed to observe 
a spin precession about cosmic ALP dark matter wind~\cite{Ruoso:2015ytk}.
Finally, a dish antenna immersed in a magnetic field may also be exploited to do broad band ALP dark matter search in the meV mass range~\cite{Horns:2012jf}.

\begin{figure}[htbp]
\includegraphics[width=0.45\textwidth]{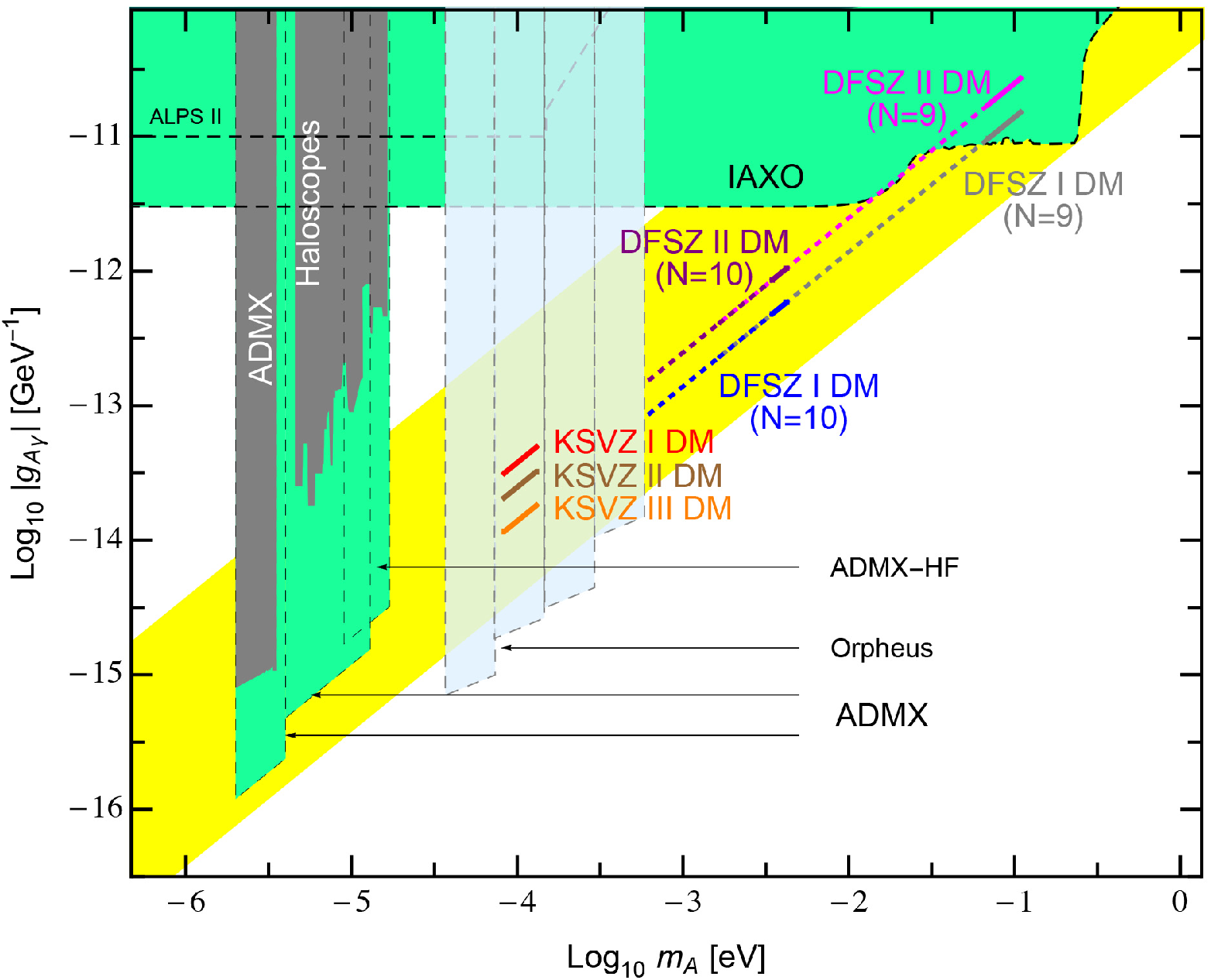}
\caption{Sensitivities to the accion-photon coupling $g_{A\gamma}$ of ongoing and planned detectors and predictions of
accion dark matter models (adapted from Ref. \cite{Hewett:2012ns}). 
The yellow region corresponds to the generic prediction for the QCD axion models.
Colored thick lines represent the parameter range in which accions explain the observed cold dark matter abundance
for KSVZ I (red), KSVZ II (brown), KSVZ III (orange), DFSZ I with $N=9$ and $|g|=1$ (gray), DFSZ I with $N=10$ and $|g|=1$ (blue), 
DFSZ II with $N=9$ and $|g|=1$ (magenta), and DFSZ II with $N=10$ and $|g|=1$ (purple).
For DFSZ models, the observed dark matter abundance can also be explained in the parameter range represented by the dotted lines
if we allow smaller values of $|g|<1$ and/or a further tuning of the phase parameter $\Delta_D$.}
\label{fig:coupling}
\end{figure}

In Fig.~\ref{fig:coupling}, we summarize sensitivities of ongoing and planned detectors and predictions of the accion dark matter models in terms of their mass $m_A$ and photon coupling $g_{A\gamma}$.
We see that the parameter region predicted by the KSVZ models can be probed by the Orpheus experiment.
Most of the parameter region predicted by the DFSZ models can also be probed by IAXO.
In particular, it covers the whole parameter region predicted by the DFSZ models with $N=9$ and $|g|=1$.
Such a parameter region is incompatible with astrophysical bounds as shown in Fig.~\ref{fig:fAmA}, and this incompatibility can be checked by IAXO in the future.
Furthermore, IAXO is expected to improve the sensitivity on $g_{A\gamma}\cdot g_{Ae}$ by two orders of magnitude compared to CAST~\cite{Barth:2013sma,Armengaud:2013LoI},
which is enough to reach the mass range $m_A\gtrsim 3\,$meV suggested by $N=10$ DFSZ models.


\section{\label{sec5} Conclusions and discussion}
In this paper, we have considered extensions of the SM in which the PQ symmetry arises as an accidental symmetry of an exact
discrete $Z_N$ symmetry. In these theories, the accion is a pseudo NG boson associated with the spontaneous 
breaking of the accidental U(1)$_{\rm PQ}$ symmetry, and it can be a candidate for cold dark matter of the Universe.
As benchmark examples of accion models, we constructed KSVZ and DFSZ like extensions of the SM by specifying charge assignments of
the exact discrete symmetry. 
We identified the lowest dimensional Planck-suppressed operator consistent with the $Z_N$ symmetry and investigated its effects on the accion potential.
It was found that CP violating semi-classical gravity effects are well suppressed if the order of the discrete symmetry is as large as $N\ge 9$.
Assuming that the PQ symmetry is broken after inflation, we also estimated the relic accion dark matter abundance by taking account of
the contributions from the decay of topological defects.
The domain wall problem in DFSZ models can be avoided due to
the existence of the small explicit PQ symmetry breaking term, which originates from the lowest dimensional Planck-suppressed operator.
It was shown that the order of the discrete symmetry must be $N\le10$ in order to avoid the overproduction of accion dark matter.
Combining the above observations, we concluded that the DFSZ accion models are phenomenologically viable if $N=9$ or $10$.

The value for the accion decay constant $f_A$ or its mass $m_A$ predicted by the accion dark matter models and its observational constraints
are summarized in Fig.~\ref{fig:fAmA}. We see that the $N=9$ DFSZ models are disfavored from astrophysical bounds,
while the prediction of the $N=10$ DFSZ models with a mass range $m_A\approx 3.5 \textendash 4.2\,\mathrm{meV}$
coincides with the region suggested by the observations of the white dwarf cooling.
It is notable that every accion dark matter model gives a distinctive prediction for coupling parameters as shown in Table \ref{tab:couplings} and Fig.~\ref{fig:coupling}
and that such a parameter region can be probed by future ALP detectors such as IAXO and Orpheus.
In particular, the meV range suggested by the $N=10$ DFSZ accion models will be decisively tested by IAXO.

It should be noted that the DFSZ accion models require a tuning of the phase parameter $\Delta_D$ in order to avoid the experimental bound from the NEDM.
The tuning becomes severe as the decay constant $f_A$ increases. In particular, $\Delta_D\lesssim \mathcal{O}(10^{-5}\textendash 10^{-4})$ is required in the $N=10$ DFSZ models.
Furthermore, it can be shown that the severity of the tuning cannot be relaxed even if we use a different cutoff scale $\Lambda_{\rm UV}$ in the higher dimensional operator
rather than $\Lambda_{\rm UV}=M_{\rm P}\simeq 2.435\times 10^{18}\,\mathrm{GeV}$ used in this paper [i.e., Eq.~\eqref{Planck_sup_operators}].
We also note that the predicted value for $f_A$ and the required values for $N$ and $\Delta_D$ would change if we fix $|g|=1$ and vary the cutoff scale $\Lambda_{\rm UV}$.

In addition to the subtle tuning of the parameter, there is another hurdle originated in the assumption of the discrete symmetry.
It has been conjectured that the exact discrete symmetry should be gauged since otherwise it can also be spoiled by gravitational effects.\footnote{This statement is less robust in comparison with that pointing to the absence of continuous global symmetries~\cite{Banks:2010zn}.}
Such a gauged discrete symmetry can arise as a low energy remnant of some continuous gauge symmetry~\cite{Krauss:1988zc}.
The problem here is that the gauged $Z_N$ symmetry (and hence its primordial gauge symmetry)
should have a color anomaly to induce the axion potential [Eq.~\eqref{V_eff_accions}], which leads to an inconsistent quantum field theory since all gauge symmetries must be anomaly free.
We note that it is still possible to avoid this inconsistency by using a Green-Schwarz type anomaly cancellation mechanism discussed in Ref.~\cite{Babu:2002ic}.
Alternatively, we can impose other continuous gauge symmetries to suppress the PQ symmetry breaking operators~\cite{Barr:1992qq,Kamionkowski:1992mf,Holman:1992us,Babu:1992cu,Cheung:2010hk}
by introducing extra particle content in addition to the minimal axion models.
In the context of string compactification~\cite{Choi:2006qj,Choi:2009jt},
both cases may happen since there exist both discrete and continuous gauge symmetries as well as numerous extra fields.

The DFSZ like extensions of the SM -- being based on a 2 Higgs Doublet Model (2HDM) --
can be probed also at the LHC. Intriguingly,  ATLAS and CMS reported hints of a new resonance in the diphoton invariant mass distribution at around 750 GeV~\cite{CERN:diphotonexcess}. 
Interpreting the latter in terms of the heavy neutral CP-even or CP-odd Higgs particles (or a superposition of the two), 
it appears that additional charged particles (e.g., vector-like quarks and leptons) contributing to the loop-induced production and decay processes 
are required to match the strong diphoton signal. 
Such an extension of the minimal 2HDM might be accommodated in a combination of the KSVZ and DFSZ accion
models.\footnote{A combination of KSVZ and DFSZ models has been considered in Ref.~\cite{Alves:2015vob} to study diboson signals from the decay of a heavy Higgs boson.}
The consequences for accion cosmology can then be worked out by following a similar 
flow of arguments as developed in this paper. 

Finally, it appears that top-down motivated orbifold compactifications of the heterotic string predict 
hidden complex scalars, vector-like exotic particles, and large discrete symmetries -- $R$ symmetries from the broken SO(6) symmetry of the
compactified space and stringy symmetries from the joining and splitting of strings --  able to give rise
to accidental global PQ-like U(1) symmetries \cite{Choi:2006qj,Choi:2009jt}. Here, the challenge is to find models where the VEV of the accidental
PQ symmetries are naturally in the range of $\sim 10^{9}\textendash 10^{10}$\,GeV rather than in the range of the heterotic string scale, $M_s =(\alpha_{\rm YM}/4\pi)^{1/2} M_{\rm P} \simeq 1.4\times 10^{17}\,\mathrm{GeV}$, where $\alpha_{\rm YM}\simeq 1/24$ is the unified value for gauge coupling parameters.
Furthermore, the heterotic string scale might appear as a ultraviolet cutoff scale $\Lambda_{\rm UV}$ in the higher dimensional operator $g\mathcal{O}_D/\Lambda_{\rm UV}^{D-4}$.
If we use $\Lambda_{\rm UV}=M_s$ rather than $\Lambda_{\rm UV}=M_{\rm P}$, the phenomenologically viable value for the order of the discrete symmetry becomes $N = 10$ or $11$,
and the predicted values for $f_A$ are modified as follows:
For $N = 10$ and $|g| = 1$, we obtain $f_A \approx (1.2 \textendash 1.7) \times 10^8\,\mathrm{GeV}$ [or $m_A \approx (3.5 \textendash 4.9) \times 10^{-2}\,\mathrm{eV}$] with a tuning of $\Delta_D \lesssim \mathcal{O}(10^{-2} \textendash 10^{-1})$,
while for $N = 11$ and $|g| = 1$, we obtain $f_A \approx (1.6 \textendash 1.9) \times 10^9\,\mathrm{GeV}$ [or $m_A \approx (3.2 \textendash 3.8) \times 10^{-3}\,\mathrm{eV}$] with a tuning of $\Delta_D \lesssim \mathcal{O}(10^{-5})$.

\begin{acknowledgments}
A.~R.~acknowledges fruitful discussions with Alex G. Dias, Maurizio Giannotti, Christophe Grojean, Igor Irastorza, Javier Redondo, and Georg Weiglein. 
K.~S.~thanks Satoshi Shirai for useful discussions.
K.~S.~is supported by the Japan Society for the Promotion of Science through research fellowships.
\end{acknowledgments}


\end{document}